\documentclass[5p]{elsarticle}
\usepackage[utf8]{inputenc}
\usepackage{lineno,hyperref}
\modulolinenumbers[5]
\usepackage{relsize}
\usepackage{booktabs}
\usepackage{color, colortbl}
\usepackage{multirow}
\usepackage{svg}

\usepackage{threeparttable}
\begin{document}

\title{\textbf{LRDB}: \textbf{L}STM \textbf{R}aw data \textbf{D}NA \textbf{B}ase-caller based on long-short term models in an active learning environment}

\author[label1]{Ahmad Rezaei}
\address[label1]{Department of Computer Science and Automation, Technische Universit{\"a}t Ilmenau, Ilmenau, Germany}
\author[label2]{Mahdi Taheri}
\address[label2]{Tallinn University of Technology, Tallinn, Estonia}
\author[label3]{Ali Mahani\corref{creff1}}
\address[label3]{Reliable \& Smart Systems Laboratory, Shahid Bahonar University, Kerman, Iran}
\author[label4]{Sebastian Magierowski}
\address[label4]{Department of Electrical Engineering and Computer Science, York University, Toronto, Canada}
\cortext[creff1]{Corresponding author}

\begin{frontmatter}
\begin{abstract}
The first important step in extracting DNA characters is using the output data of MinION devices in the form of electrical current signals. Various cutting-edge base callers use this data to detect the DNA characters based on the input. In this paper, we discuss several shortcomings of prior base callers in the case of time-critical applications, privacy-aware design, and the problem of catastrophic forgetting. Next, we propose the LRDB model, a lightweight open-source model for private developments with a better read-identity (0.35\% increase) for the target bacterial samples in the paper. We have limited the extent of training data and benefited from the transfer learning algorithm to make the active usage of the LRDB viable in critical applications. Henceforth, less training time for adapting to new DNA samples (in our case, Bacterial samples) is needed. Furthermore, LRDB can be modified concerning the user constraints as the results show a negligible accuracy loss in case of using fewer parameters. We have also assessed the noise-tolerance property, which offers about a 1.439\% decline in accuracy for a 15dB noise injection, and the performance metrics show that the model executes in a medium speed range compared with current cutting-edge models.
\end{abstract}


\end{frontmatter}


\section{Introduction}

Based on the production of a specific ionic current for each nucleotide, the concept of a Nanopore device was first described in 1989 for DNA sequencing purposes. This led to the development of the actual device in 2001 \cite{howorka2001sequence}. This device contains several pores with a stable current along the passage and sensors that observe the current value through time. Each of these pores is designed for passing DNA strands inside them, affecting the ion current along their passage. This current is used as the sensory data for further identification of various nucleotides forming a strand of DNA\cite{taheri2021high, taheridevelopment}
. Several pores are assembled in a small Min-ION device to speed up this procedure, making it portable for remote use and a good candidate for time-critical genomic applications.  
On average, 450 base pairs of DNA strands are passed through the pore each second (bp/s), and the current is sampled at 4 kHz frequency \cite{nan_spd}. This means 1 K-mers (sub-strings) should be identified in each second for each DNA character. To further translate these K-mers to corresponding DNA characters, several methods and applications such as MinKNOW are proposed under the concept of base calling. The main objective of base callers is to translate this raw electric current signal to a DNA sequence, which is done through a sophisticated procedure.

Until recently, there has been rapid competition between various base calling approaches demonstrating their performance's significance and determination. This trend has evolved to perform with higher base-pair quality as an output of its primary objective. Base callers are mainly divided into Event and Raw Database callers. In the prior one, several preprocessing steps and transformations are applied that slow down the overall process, while in the latter, raw current signals are directly used in the base calling that enhances the base calling run time manifold\cite{taheri2021novel}.

Although MinION technology had paved the way for high-speed DNA sequencing applications, the methods first used in transforming the current outputs of the MinION devices did not compete with the expectations as they were using event data. These models lacked the property to operate parallel with the MinION devices because their lower speed stemmed from processing Event data. They needed several steps in transforming the event data into solvable data for the proposed methods, e.g., DeepNano \cite{bovza2017deepnano} used a time-consuming segmentation or Nanocall \cite{david2017nanocall} benefited from Hidden Markov Models which were impractical in modeling long sequence dependencies.

However, a new approach was first explored in Scrappie \cite{magi2018nanopore} that eases the procedure by base calling DNA reads directly from the raw signal output of MinION. Utilizing raw DNA omits several preprocessing stages and reduces overall time. Hence, due to their improvement in comparison with prior versions of base callers, raw base callers superseded former methods.
In the meantime, several raw base callers were proposed to reach a higher accuracy e.g. BasecRAWller \cite{stoiber2017basecrawller}, Chiron \cite{teng2018chiron}, Guppy \cite{wick2019performance}, SACall \cite{huang2020sacall}, Causalcall \cite{zeng2020causalcall}, CATCaller \cite{lv2020end}, and Halcyon \cite{konishi2021halcyon} (their properties are addressed in section \ref{sec:prb}). 
Mentioned base callers have generally improved the base calling; however, several shortcomings remain. As discussed before, base calling is a time-critical application and training the model from scratch usually takes several epochs to converge. Also, the dynamic nature of the bacterial genome questions the generality of the models as the base callers only report their base calling performance for a sub-category of bacteria (Klebsiella pneumonia species). Hence, in LRDB, we aim to use two concepts of Transfer learning \cite{torrey2010transfer} and Fine-tuning \cite{howard2018universal} to expand the generality for already trained model weights, which reduces the training time dramatically. Also, we provide precise details of our open-source model for readers, which helps with the privacy issues that may arise using the private models for base callings. This provides a reproducible base caller that the users can create their models with various options for model parameters and model performance. Furthermore, we aim to propose LRDB, which achieves higher accuracy than existing models for target bacteria species. Additionally, the performance is comparable to cutting-edge ONT base callers.

The rest of the paper is contributed as follows:

\begin{enumerate}
    \item The state-of-art about raw base callers are elaborated to discuss the drawbacks of former methods and the characteristics that can be improved.
    \item A brief discussion about Long-Short-Term-Memory and composite auto-encoders is provided, which are used in the LRDB model.
    \item We develop our LRDB model under various constraints, and a discussion on the different aspects of our network's properties is provided.
    \item The optimal model is selected and used in several case studies.
    \item Eventually, our research results are addressed, and the paper is concluded.
\end{enumerate}

\section{Prior Raw Basecallers}
\label{sec:prb}
In this section, we emphasize different raw base calling approaches and the most notable improvements in recent years. 

As discussed before, various Deep Learning models were developed to directly process raw current signals of Nanopore devices into DNA sequential reads. The pace of most of these base callers (ranging from less than 1000 bp/s for Halcyon to 7500 bp/s for Bonito base caller) keeps up with these devices and can be executed concurrently with Nanopore devices. Previously, the first raw base caller model was developed named BasecRAWller \cite{stoiber2017basecrawller} that uses raw signals as inputs and performs segmentation procedures inside the model. Although BasecRAWller questioned the efficacy of raw base calling on producing accurate DNA reads, it only reached a read identity (explained in section \ref{sec:perf_metrics}) of 81\% for most experimental cases and was still dependent on segmentation procedure.

In the meantime, this question of raw data usage or dependence on Event-simulated data was investigated by several researchers, and its conclusion proved the fact that raw data utilization in base callers enhances the quality, accuracy, and yield of base calling.

Chiron \cite{teng2018chiron} is developed as the first base caller fully independent from Event-simulated data. Chiron base calls DNA reads directly from raw input signals, and its network consists of three parts; convolutional neural networks (CNN), recurrent neural networks (RNN), and connectionist temporal classification (CTC). It demonstrates a proper read identity in the output; however, it is regarded as the slowest raw base caller model on a CPU with a high number of parameters (approximately 2.6 million).

In the meantime, ONT company has developed several raw base caller models, with regular updates on each base caller. 

For instance, Albacore is developed up to version 2.3.4, which uses a flip-flop algorithm in its latest versions, and also, Guppy which performs similarly to Albacore, is developed to use GPU acceleration. Additionally, with the release of the newest pore version, ONT has released a base caller called Flappy, which performs base calling based on CPU. Overall, ONT base callers have a similar read identity around 90\%, which is a significant improvement compared to former methods. Lack of proper method to train these base callers leads to their long training times, e.g., Guppy needs 36.5 - 48 hours to be trained in \cite{wick2019performance} with NVIDIA P100 GPU six core (12 thread) Intel Xeon W-2135 CPU, 32 GB RAM.

Furthermore, SACall base caller was proposed in \cite{huang2019attention}. This model yields better results than ONT base callers with a 1\% accuracy improvement and is publicly available, allowing users to train their own models. The model consists of convolutional layers, transformer self-attention layers, and a CTC decoder in the end.

Causalcall \cite{zeng2020causalcall} is another raw base caller based on Temporal Convolutional Networks (TCN). This base caller focuses on the quality of base calling by reducing the error rates in the base calling procedure. However, the reported read identity is below 90\% accuracy, which shows that the aforementioned base callers like ONT base callers or SACall are more suitable for their read identity properties. Also, one major drawback of this work is the long training time for the model to converge, as the authors reported that three days of training is needed for the convergence on NVIDIA1080ti GPU with 12 GB memory.

CATCaller was proposed in \cite{lv2020end}, which benefits from augmented-convolutional transformers in the model structure. CATCaller aims at speeding up the base calling procedure using various techniques, such as allocating the model based on float16 data structure and speed-up by GPU parallelization. Moreover, they enhance the accuracy by 0.4\% in comparison to the SACall base caller. The training procedure in this base caller is not elaborated on, and the authors do not provide exact training information.

Eventually, Halcyon \cite{konishi2021halcyon} was proposed as an open-source base caller. The accuracy results are competitive to that of ONT base callers, while it is the second slowest raw base caller compared to all the existing base callers. Also, the authors have not elaborated in detail about their training phase and have used their own data for the Human genome instead of using bacterial samples such as other base callers to provide a fair comparison.

Base callers are meant to be used in time-critical and portable applications, and base calls diverse genome structures in a short time. For instance, whole DNA sequencing process (in which the base callers play a vital rule) is executed in portable laboratories in places such as Antarctica \cite{johnson2017real}, coal fields \cite{edwards2017deep}, arctic \cite{edwards2016extreme}, or international space station \cite{castro2017nanopore}.  Moreover, time is an important factor in detecting outbreaks by using the same DNA extraction and identification procedure\cite{taheri2021hardware}, which base calling is an important part of translating the physical signals into DNA characters. For example, authors in \cite{mitsuhashi2017portable} identify bacterial composition in only 2 hours, or authors in \cite{quick2015rapid} demonstrated the need for only half a day to genotype the Salmonella outbreak in a hospital. 

Several significant problems remain unsolved based on the portability and time-critical need for base calling applications. The whole DNA identification procedure is questioned if the user base calls a species with a distinct structure from training data as the base caller behaves unforeseen against the new data. In this case, the model is often trained for a closely related species, and the training data is available for the desired accuracy. For instance, in the already discussed base callers, 50 different bacterial samples are used as the training data to target a subset of a similar bacterial sample as the test data (the same test samples used in this paper in section \ref{sec:perf_metrics}). However, starting from scratch to train the base callers is time-consuming and against the time-critical nature of the base calling. So the need for an extendable base caller model which can reach the desired accuracy with just training a subset of the model parameters is highlighted. 

The structure of some former base callers, such as Guppy and Flappie, is private, and users are unaware of privacy issues that may threaten their data. Henceforth, it is desired for users to implement the model for themselves rather than using pre-trained base callers, which also highlights the time-consuming train phase more than before. The proposed base caller in this paper is explained in detail, making it suitable to be reproduced by a third party.

More importantly, the training phase time consumption is the sheer volume of data used, which also triggers a well-known problem in Deep Learning called catastrophic forgetting. We suggest an approach to only use a subset of the data and reach a reasonable accuracy for target bacteria species. 

To solve these issues, an LRDB base caller is proposed. This base caller is an end-to-end base caller classified under this raw base calling model. Moreover, it reaches a higher accuracy to target bacteria species than current base callers, even using a reasonable number of parameters. Additionally, mentioned issues are addressed and solved using LRDB in a Transfer learning and fine-tuning scenario, reducing training time by restricting it to a subset of model parameters. The noise resilience property of the LSTM caller is further reported.

\section{Preliminary knowledge about time-series models}

\subsection{Long Short Term Memory}
\begin{figure*}[ht]
    \centering
    \includegraphics[width=2\columnwidth]{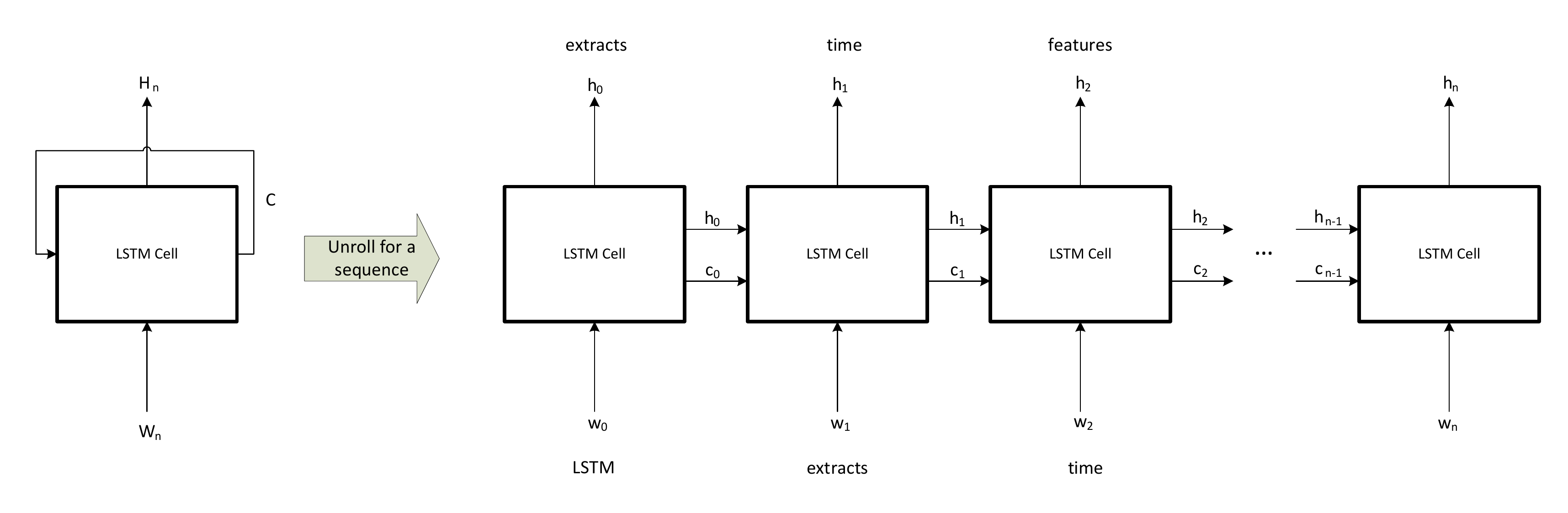} 
    \caption{LSTM structure elaborated}
    \label{fig:LSTM_cell}
\end{figure*}

Feedforward neural networks propagate the information from the input to the network's output in order to map an input vector (in the form of an image or a sequence) to the output vector. However, LSTM networks use loops to create connections between different parts of a sequence and they are applied to sequence-to-sequence mapping problems \cite{greff2016lstm}. Fig. \ref{fig:LSTM_cell} is a simplified illustration of sequence-to-sequence mapping for predicting the next word in a sentence. $W_k$ and $H_k$ stood for the input sequence and predicted output, and $C$ is the time-dependent weight that preserves the network's timing features.

In more details, if we unroll a LSTM unit in time, at each time step $t_k$, reads $w_{t}$ as an input. It takes also $c_{t-1}$ and $h_{t-1}$ as the previous hidden state and previous memory respectively in order to compute its hidden state ($h_{t}$). This hidden state is further passed into a Softmax function to predict the probability of the next word across the vocabulary.

\begin{figure}[ht]
    \centering
    \includegraphics[width=\columnwidth]{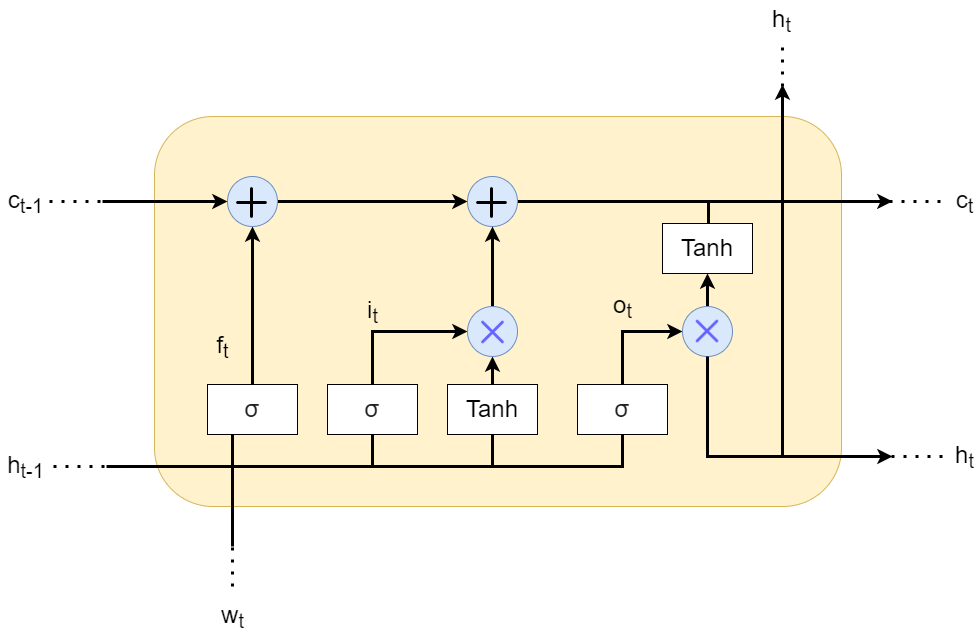}
    \caption{LSTM cell internal structure}
    \label{fig:int_cell}
\end{figure}

The essential element of an LSTM is the so-called short-term memory cell, which stores the information over time (see Fig. \ref{fig:int_cell}). The information in the memory cell is refreshed each time step by forgetting a portion of old information and accepting new relevant information. For this sake, LSTM uses a forget gate which takes $w_{t}$ and $h_{t-1}$ as inputs and outputs a value of $f_t$, which shows how much of the information from $h_{t-1}$ should be preserved. This output value is between 0 (translating to forgetting all the past data) and 1 (preserving all the past data). Next, the new information from $w_{t}$ is added to the memory cell by using an input gate $i_t$, and this information is further used in computing the value of $h_t$. The output gate $o_t$ also uses $w_t$ and $h_{t-1}$ to produce the output for each time step.

This architecture enables the LSTMs to capture long-term dependencies in a sequence. For this reason, LSTM-based models are used in various applications ranging from speech recognition, video analysis, and language modeling. LSTMs are applied to areas of knowledge where we have little or no information about meaningful time-dependent features present in a sequence and they have shown significant improvements in these feature extraction properties. In our case, we have benefited from them in base-calling long sequences of the current signal as the input to grasp the unknown dependencies in these signals, which map into a DNA character.

\subsection{Composite Auto-encoder}
Auto-encoder networks have a wide range of applications and are actively applied for various use cases \cite{pu2016variational}. The most common use cases of Auto-encoders are noise removal from the data, data reconstruction, and lightweight capturing of meaningful features.

The architecture of an Auto-encoder network consists of two sequentially connected parts which encode and decode the information to build the input data in the output. Both of these parts are trained in an unsupervised manner to reach a low regression loss, and then the decoder part is removed from the network, and the encoder part, which now contains necessary trained weights based on the input data, can be used in Deep Learning models for higher noise resistance and more proper training.

Although the variational Auto-encoders are primarily used in convolutional networks, more specific Auto-encoders exist for LSTM networks to capture the time-dependent features in a sequence of input. These Auto-encoders are trained using the same unsupervised regression phase, and their encoder part can be used in predictive models.

In this paper, we follow the approach investigated in \cite{srivastava2015unsupervised} to implement an Auto-encoder (see Fig. \ref{fig:auto_comp}), which captures the time-dependent features for reconstructing the electrical current input signal at the same time step (Current Signal (t0)) and in one time-step ahead (Current Signal (t1)). This Auto-encoder is called "Composite Auto-encoder" because it has one encoder part and two decoder parts each responding to the prediction of the input signal in current and next time steps. This Composite Auto-encoder improves accuracy on various Deep Learning models, making it suitable for our predictive model. Different model parameters for the Composite Auto-encoder are evaluated in the training stage to reach the minimum possible regression loss. After the training phase, we remove the decoder and use the encoder part as the first layer in our LRDB model. 

\section{Proposed Model}
In this section, we first define the structure of our proposed model. Next, we elaborate on various characteristics of the LRDB model. 

\subsection{LRDB Model}
The overall structure of our proposed model consists of two parts, respectively, for encoding the information from our time-series input signals and decoding this information to their actual location on the DNA strand in the output. In the following subsections, we first elaborate on these parts and explain the overall assembled position of the model.

\subsubsection{Encoder model}

\begin{figure}[ht]
    \centering
    \includegraphics[width=0.4\columnwidth]{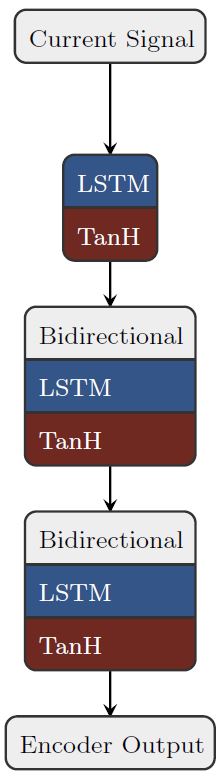}
    \caption{Encoder part of LRDB base caller}
    \label{fig:Encoder}
\end{figure}

Encoder model has one input and one output tensor respectively with a size of $(batch\_size,s\_ln,1)$ and $(batch\_size,\newline s\_ln, 2n1)$, where $batch\_size$, $s\_ln$, and $n1$ consecutively stand for batch size, length of input signal, and number of units in stacked bidirectional LSTM layers. In the encoder part, the input signal is first fed into a single LSTM layer with $n0$ number of units, and then the outputs of this LSTM layer are fed into two other stacked bidirectional LSTM layers to produce the outputs. The overall architecture of our model is depicted in Fig. \ref{fig:Encoder}. 

This architecture's first one-directional LSTM layer extracts the most meaningful information from input signals and provides subsequent layers with primary information tensors. Therefore, in our design, we benefit from a Composite Auto-encoder model to extract the most prominent weights from input signal series and initialize our model with these weights instead of applying random initialization techniques.

\begin{figure}
    \centering
    \includegraphics[width=0.8\columnwidth]{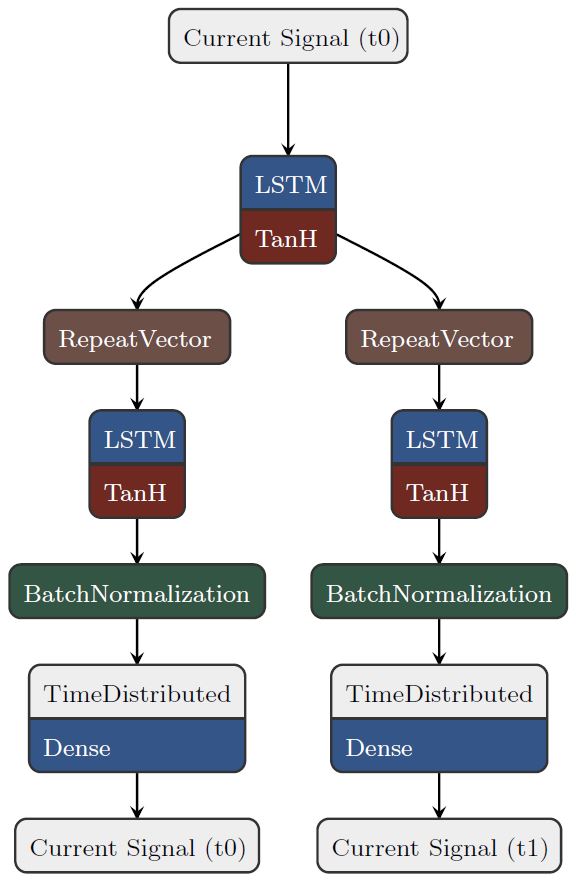}
    \caption{Composite Auto-encoder architecture used in training first one-directional LSTM of LRDB}
    \label{fig:auto_comp}
\end{figure}

The composite Auto-encoder model (shown in Fig. \ref{fig:auto_comp}) desirable for our LSTM layer consists of one primary LSTM layer with the same number of units as our Encoder model and two parts which solve different regression problems beneficial for meaningful weight extraction. The first part is trained to predict the Current Signal (t0) at the same time step as the input signal, and another is trained to predict the Current Signal (t1) at the next step. Both of these parts use the same architecture, of 200 units of LSTM layers followed by Dense layers for input prediction.

Using Composite auto-encoders has several beneficial impacts on our model: The vanishing gradient problem, which can hinder our training, is dealt with. As updating weights of this layer requires the calculation of gradients in the outer layers of the model, and based on characteristics of LSTMs, which only use Tanh and Sigmoid activation functions, gradients converge to small numbers close to zero leading to improper updates. Whereas initializing this layer with trained weights from our composite auto-encoder model diminishes the need for more significant weight updates. Composite auto-encoders are very helpful for extracting temporal features of the input. This model can also improve the model's noise invariance factors if it is also trained on noisy data.

So we can also leverage noise impacts always present in measurements and electronic devices, including this part.

Encoder parameters, $n0$, $n1$, and $s\_ln$ are defined during training, and after this stage, the outputs are fed to the decoder part, which analyses the information to predict the DNA reads' values. In the following paragraphs, we define the decoder structure of our model.

\subsubsection{Decoder model}

\begin{figure}[ht]
    \centering
    \includegraphics[width=0.8\columnwidth]{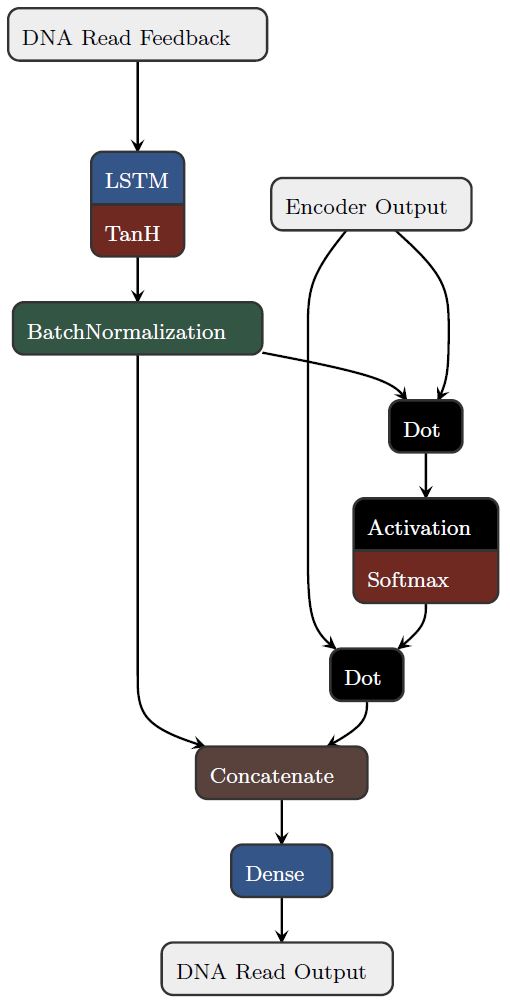}
    \caption{Decoder part of LRDB base caller}
    \label{fig:Decoder}
\end{figure}

Here, we explain the decoder architecture depicted in Fig. \ref{fig:Decoder}. The decoder part contains three inputs and one predicted output (DNA Read Output). The first input (DNA Read Feedback) is fed into states of the one-directional LSTM layer in the decoder part, and the second input (Encoder Output) is fed into inputs of Dot matrices layers in the attention part after the LSTM layer. The shape of the first input is $(s\_ln, 2n1)$, and the other inputs are the outputs of the encoder part.

DNA Read Feedback is the predicted output of the earlier timestamps added to the decoder as an input. The decoder performs LSTM-based computation on the time-series predicted output, and then it will compute the attention values and feed the classifier in the model's output to predict DNA reads.

Both of the mentioned parts are assembled to form the LRDB model. In other words, the encoder part computes time-dependent matrix values, and next, the decoder repeats 40 times and, using the attention mechanism, predicts 40 (used frequently in the literature) possible DNA characters. In the speed evaluation of our model, we first report the speed of running the whole model as a single object, then we will run the two different parts of it separately and report the possible speed-up caused by running the encoder part only once.

\subsection{Model characteristics}

Several beneficial factors are highlighted in this section, which are further reported and discussed in the following sections.

\subsubsection{Constraint on train data volume}
Based on three discussions, the volume of data used in the training phase should be reduced.

First, the training time in such time-critical scenarios is essential; For instance, we need to start the training from the beginning, each time we include a new Bacterial species in former base callers, or we should wait long-training time to train on the whole data set based on the lower performance of mobile devices. Furthermore, the dynamic behavior of the base callers allows the users to adapt their base caller faster to new DNA samples in a time-critical scenario, which the base caller should perform in a short matter of time. Henceforth, we aim to reduce the training time and make it more feasible for users to train the LRDB model for their samples to have higher efficiency. In our training scenario, less training data is used to train the network than in other models. For example, the extent of data used for training in \cite{wick2019performance} is 108 Gigabytes, which makes the training time-consuming. This time consumption is based on the fact that the network needs to be trained for several epochs on the same data, and in this paper, we aim to reduce the train phase time.

Second, because of privacy issues, training private base callers is of value, and using less data in the training phase enables the user to train and use their base callers. Current base callers such as Guppy are private, and the information about their structure is not available to users; so the users can not either train them based on their needs or make sure about the safety and protection of their data in the base calling process.

Also, exceeding the volume of data in the training phase leads to a known issue called "Forgetting catastrophe." This issue stems from the network's ability to hold its former learned connections during the training in the presence of a considerable amount of training data. This issue is discussed, and several state-of-art methods are proposed to solve it, but the problem still exists in the naive training phase without taking the necessary measures.

This issue is not discussed in the training of the prior models, and no insights are offered on how to deal with this phenomenon. In our paper, we aim to bypass this problem by using fewer data in the training procedure to reach the desired generalization of the trained model. Moreover, we benefit from transfer learning and fine-tuning to enhance this generalization. The difference between our approach and other base caller's approach is that LRDB is trained for specific bacterial samples or similar bacterial genomes to reach a higher read identity, enhanced time-critical usage, and lower training time, whereas other base callers train only one model for all the existing bacterial samples, which leads to base callers with more deficient read identity and no solution for convenient training time.

\subsubsection{Base caller extendability}
Following our discussion on our training data selection, it should be highlighted that the bacterial genome structure is diverse \cite{casjens1998diverse}. Thus, in the case of training a general base caller for all bacterial samples, the users do not have the best accuracy performance.

For such an issue, the prior base callers lag as they should be trained for new bacterial samples from the starting point if the accuracy drops. Furthermore, transfer learning is not investigated for these prior base callers. However, LRDB has proven helpful by being trained for a specific range of closely related bacterial samples or a few bacterial groups and allowing us to transfer this learned knowledge into a base caller for another type of bacterial samples. This transfer learning step reduces the training time of the base caller by only performing fine-tuning on the classification layer to achieve a reasonable accuracy and, if desired, extend the fine-tuning to prior layers in the model to reach the highest possible accuracy for the base caller.

In conclusion, we benefit from transfer learning and fine-tuning to reduce the training time for new samples to achieve the best accuracy results for diverse bacterial genomes.

\subsubsection{Composite Auto-encoder}
Our model benefits from a composite auto-encoder that extracts more time-dependent features from the input current signals, leading to higher accuracy in predicting the DNA sequence. We elaborate on this property of our model by comparing the same model without an added auto-encoder layer in the beginning.

\begin{figure*}[t]
    \centering
    \includegraphics[width=0.7\textwidth]{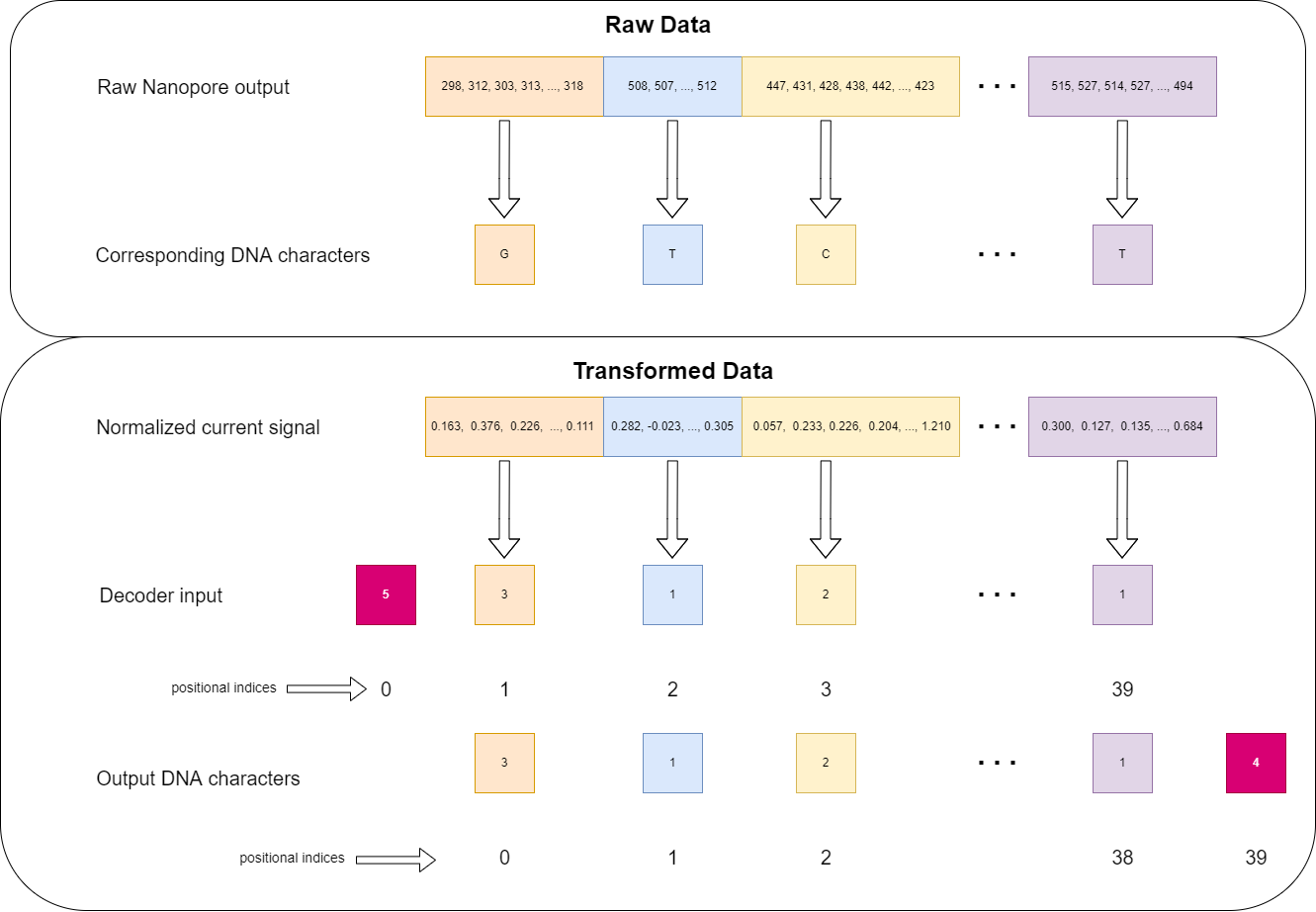}
    \caption{An illustration of raw input and output data with their respective transformed data for the training stage. The starting point and end point markings are colored red.}
    \label{fig:elicitation_data}
\end{figure*}

\subsubsection{Parallel operation of encoder and decoder parts}
In this model, the encoder and decoder parts are merged only in the attention mechanism, and the decoder does not take hidden layers of the encoder as its initial hidden states. This will increase the model's speed as the decoder and encoder can operate in parallel before feeding to the inputs of the attention mechanism. In more detail, for the first iteration (out of 40 iterations), the model is executed, and the final values of the encoder are stored; this first execution can be performed on the whole model, or the encoder and decoder can be executed in parallel (due to their independence) and passed through the attention mechanism. In the rest of the iterations (39 left iterations), the encoder part can be removed from the model, and the decoder part is executed with the stored data of the encoder from its first iteration being fed into the attention mechanism.

\section{Model characteristics investigation}
\label{sec:train_prc}
This section explains the training phase to reach our desired accuracy. Various parts of this phase are elaborated in different sub-sections.

During the training and evaluation of the LRDB, we have used the data publicly available at \cite{Wick2019_train}. We have used Tensorflow 2.3 platform on Tesla P100-PCI-E GPU 16Gb with 25 Gbyte RAM during all the experiments.

\subsection{Input data processing}

Training data in the LRDB, the current signal, and DNA reads are cropped into 300 and 40-length sequences such as former methods \cite{wick2019performance}.

The current signal is cropped from the raw current signal of the Min-ION device. The current strings are overlapped with a 20 signal length to cover all currents related to a DNA character, leading to a more precise training with no missing parts. The current signals are further normalized with a mean of zero and a standard deviation of one, which enables our predictive model to have a more smooth training curve.

The DNA reads are mapped into single integer numbers ranging from 0 to 3. The end of each DNA read is marked and padded with the number 4 to reach the 40 sequence length; this enables us to predict variable lengths of DNA reads. Next, as the model contains a feedback part in its attention mechanism, we use a constant number 5 integer value as the starting point marking for the first time step.

For better clarity, an example of these transformations is shown in Fig. \ref{fig:elicitation_data}.

\subsection{Optimizer selection and Hyper-parameters}

The model was trained with an early stopping point, where the accuracy level remained unchanged for three epochs or reduced. At the end of the training procedure, the highest accuracy is saved as the final checkpoint. We have derived these numbers based on experimental trials to achieve the best result and avoid unnecessary training iterations, which may cause the over-fitting issue.

During the training procedure, we examined various optimizers such as Adam, Nadam, and SGD for our training. The observations show that LRDB with the same model parameters behave differently based on its optimizer, e.g., the model reached a relatively higher accuracy using Nadam optimizer with the same hyperparameters. The outcome of our experiments selecting the optimizer algorithm is shown in Table. \ref{tab:opt_sel}.

\begin{table}[ht]
    \centering
    \caption{Various optimizers are evaluated for LRDB model with "Salmonella enterica 201006152" bacteria samples}
    \label{tab:opt_sel}
    \begin{tabular}{lcc}
    \toprule
        Optimizer & Final Accuracy & Epochs \\
        \midrule
        Nadam & 94.85 & 4\\
        Adam & 94.62 & 4\\
        SGD & 73.16 & 10\\
        Adamax & 94.53 & 5\\
        Adadelta & 65.44 & 5\\
        \bottomrule
    \end{tabular}
\end{table}

As demonstrated in table \ref{tab:opt_sel}, the Nadam algorithm was chosen for the model training, and the hyperparameters given in the table \ref{tab:hyp} were selected by trial to be used for training the LRDB.

\begin{table}[ht]
    \centering
    \caption{Parameters used in this paper for LRDB model}
    \label{tab:hyp}
    \begin{tabular}{lc}
    \toprule
        Parameter & Value \\
        \midrule
        batch size & 64\\
        max epochs & 10\\
        base\_lr & 0.001 \\
        momentum & 0.9 \\
        weight decay & 0.0005\\
        optimizer & Nadam\\
        lr\_policy & Exponential\\
        early stopping patience & 3\\
        \bottomrule
    \end{tabular}
\end{table}

\begin{table*}
    \centering
    \caption{training performed on results of the MinION system on "Salmonella enterica 201006152" bacteria samples for various model parameters.}
    \label{tab:model_params}
    \begin{tabular}{cccccc}
        \toprule
        Auto-encoder & Encoder layer 1 & Encoder layer 2 & Decoder & Total parameters & Final accuracy \\
        \midrule
        - & 25 & 25 & 50 & 31,505 & 88.95 \\ 
        - & 50 & 50 & 100 & 123,005 & 92.93 \\ 
        - & 70 & 70 & 140 & 239,405 & 94.22 \\ 
        - & 100 & 100 & 200 & 486,005 & 94.78 \\
        \midrule
        100 & 25 & 25 & 50 & 92,105  & 91.40 \\ 
        100 & 50 & 50 & 100 & 203,405 &  93.62\\ 
        100 & 70 & 70 & 140 & 335,645 &  94.50\\ 
        100 & 100 & 100 & 200 & 606,005 & 94.38 \\ \midrule
        200 & 25 & 25 & 50 & 232,905 & 93.14 \\ 
        200 & 50 & 50 & 100 & 364,205 & 94.02 \\ 
        \rowcolor{lightgray} 200 & 70 & 70 & 140 & 512,445 & 94.95 \\ 
        200 & 100 & 100 & 200 & 806,805 & 94.85 \\ \bottomrule
    \end{tabular}
\end{table*}

\subsection{Model parameters}

LRDB consists of three complex parts with a variable number of units. The selection of model parameters in each of these parts affects the final achieved accuracy of the model. Hence, we first train the auto-encoder with various parameters, and next, we train various LRDBs with and without auto-encoders and with different parameters in each part.

To evaluate the model parameters, "Salmonella enterica 201006152" Bactria species are randomly selected to perform the training. As the data, optimizers, and hyperparameters are the same during the training procedure, the model's behavior is evaluated with a fair judgment.

\subsubsection{Training Composite Auto-encoder}
\label{sec:tca}
Various LSTM units for auto-encoders are used to evaluate the Composite Auto-encoder for extracting time features of our signal sequences. We benefited from the Adam optimizer and Mean Squared Error Loss (explained in equation \ref{eq:eq_loss}) during the training phase and trained the models for ten epochs on "Salmonella enterica 201006152" signals. The Composite auto-encoder performs a regression to predict the input signals at the same timestamp and the next time stamps with the following loss function formed from both of these predictions.

\begin{equation}
    L_{MSE}(y,\hat y)=L^t_{MSE}(y,\hat y) + L^{t+1}_{MSE}(y,\hat y)
    \label{eq:eq_loss}
\end{equation}

$L^t_{MSE}(y,\hat y)$ in equation \ref{eq:eq_loss} stands for loss function computation for prediction of the same time frame output, and $L^{t+1}_{MSE}(y,\hat y)$ is the same loss function calculated for predicting one time frame ahead of the input (the second output of the Composite auto-encoder).

The results are depicted in Table \ref{tab:enc_params}, which clearly shows that the loss function for LSTM auto-encoder with 200 units has reached its minimum compared to others. In this Table, we have also included the total parameters of the composite auto-encoder and the parameters in the LSTM layer (which is used in the LRDB later). In our experiment, we have only covered Composite auto-encoders with less than 200 LSTM units, which are used frequently in the literature. Having more LSTM units lead to an excessive number of parameters in the final LRDB model and slower model performance simply because of more having more operations. LSTM part of auto-encoders with 100 and 200 units are chosen to scrutiny the impact of auto-encoders on our proposed model. The prior encoder has fewer parameters in its LSTM layer (40,800 parameters) and extracts fewer features than the latter based on the higher regression loss reported in the Table. The latter has the lowest Loss function value (0.064) throughout the Table and is potentially more suitable for further usage.

\begin{table}
    \centering
    \caption{training performed on results of the MinION system on "Salmonella enterica 201006152" bacteria samples for various model parameters.}
    \label{tab:enc_params}
    \begin{tabular}{p{1.7cm}p{2cm}p{2cm}p{1cm}}
        \toprule
        LSTM layer units & Model parameters & LSTM layer parameters & Final Loss \\
        \midrule
        50 & 50,902 & 10,400 & 0.366  \\ 
        100 & 201,802 & 40,800 & 0.249 \\ 
        125 & 314,752 & 63,500 & 0.219 \\ 
        150 & 452,702 & 91,200 & 0.144 \\ 
        175 & 615,652 & 123,900 & 0.118 \\ 
        200 & 803,602 & 161,600 & 0.064 \\ 
        \bottomrule
    \end{tabular}

\end{table}

\subsubsection{Evaluation of model parameters}
\label{sec:param_setting}
Various LRDB models are trained and evaluated to select suitable parameters for our studies. We have used two variations of LRDBs in total, with or without the Composite auto-encoder part, to observe the impact of the encoder on the training phase. The LRDB, as shown in Table. \ref{tab:model_params} is trained for our selected 100 and 200 unit LSTM parts of auto-encoders. For each of these three variations, we have trained the model for 25, 50, 70, and 100 unit bidirectional encoders and their corresponding decoder to have the model running.

As observable in the Table, the composite auto-encoder extracts more useful time-dependent features for the prediction. The Table explicitly shows that using 200 unit Auto-encoder leads to higher accuracy results as expected, and utilizing 100 unit Auto-encoder part leads to partial improvement in accuracy results in some cases. Hence, the results shown in this Table agree with the discussion in section \ref{sec:tca} that the auto-encoder with 200 LSTM units achieves a higher prediction accuracy and, as expected from its lower loss value in the training phase of the Composite auto-encoder (Table. \ref{tab:enc_params}). 

\begin{table*}[ht]
    \centering
    \caption{Bacterial Data used during the training phase}
    \label{tab:bacteria samples}
    \begin{tabular}{lcccc}
    \toprule
        Species Name & Domain & Phylum & Class & Order \\ \midrule
        Salmonella enterica 201006152 & Bacteria & Proteobacteria & Gammaproteobacteria & Enterobacterales \\ 
        Comamonas kerstersii MSB1 7G & Bacteria & Proteobacteria & Betaproteobacteria & Burkholderiales\\ 
        Acinetobacter nosocomialis MINF 5C & Bacteria & Proteobacteria & Gammaproteobacteria & Pseudomonadales
        \\ 
         Klebsiella pneumoniae NUH27 & Bacteria & Proteobacteria & Gammaproteobacteria & Enterobacterales
        \\ 
        Klebsiella pneumoniae KSB1 6G & Bacteria & Proteobacteria & Gammaproteobacteria & Enterobacterales
        \\ \bottomrule
    \end{tabular}

\end{table*}

In conclusion, we have chosen the model consisting of 3 encoder layers with 200 units for the first layer followed by two 70-unit bidirectional LSTMs and one layer decoder with 140 unit unidirectional LSTM layer followed by attention and classifier (highlighted in Table. \ref{tab:model_params}). In the section \ref{sec:res_n_discussions}, we report various characteristics of our selected model in various scenarios to confirm that desirable generalization is achieved only using a portion of training data. This model is further compared with the models mentioned earlier to point out its superiority against former proposed models.

\section{Results and Discussions}
\label{sec:res_n_discussions}

In this section, we first deal with the forgetting catastrophe issue by using a sub-set of data to train our model and achieve the desired model generalization. Next, we report read identity values for several bacterial samples and highlight the noise-tolerant characteristics of LRDB. Finally, the model performance in a real-world scenario is evaluated and discussed.

\subsection{Model training}
\subsubsection{Transfer learning and accuracy evaluation}
\label{sec:tlaae}

An LRDB model with aforementioned parameters shown in table \ref{tab:model_params} is selected from the evaluations in section \ref{sec:train_prc}.

To better deal with forgetting catastrophe and bypassing the issue, the model is trained for one bacterial sample, and after reaching the accuracy pick for this sample, it is trained with the following bacterial data. We train the complete model architecture for the first bacterial data and use a transfer learning approach for the subsequent bacterial data.

In this transfer learning approach, the trained model parameters' are frozen except for the parameters in the classifier layer. Then, the model is trained with a low learning rate (0.0005) to reach its highest accuracy point. During each epoch, the model is evaluated to check if the accuracy is increasing or not. Next, to reach its maximum accuracy, we will defreeze the model parameters in the decoder part and train the model again. After reaching maximum accuracy, more layers of encoder parts are defrozen until the model is fully trained.

This training procedure is performed on five different bacteria species, depicted in table \ref{tab:bacteria samples}, and the evaluation and inference results are reported in each stage. The evaluation results in these five stages give us valuable insight into the model's generalization, which helps select the model more accurately based on calling more bacterial samples.

\begin{figure}
    \centering
    \includegraphics[width=\columnwidth]{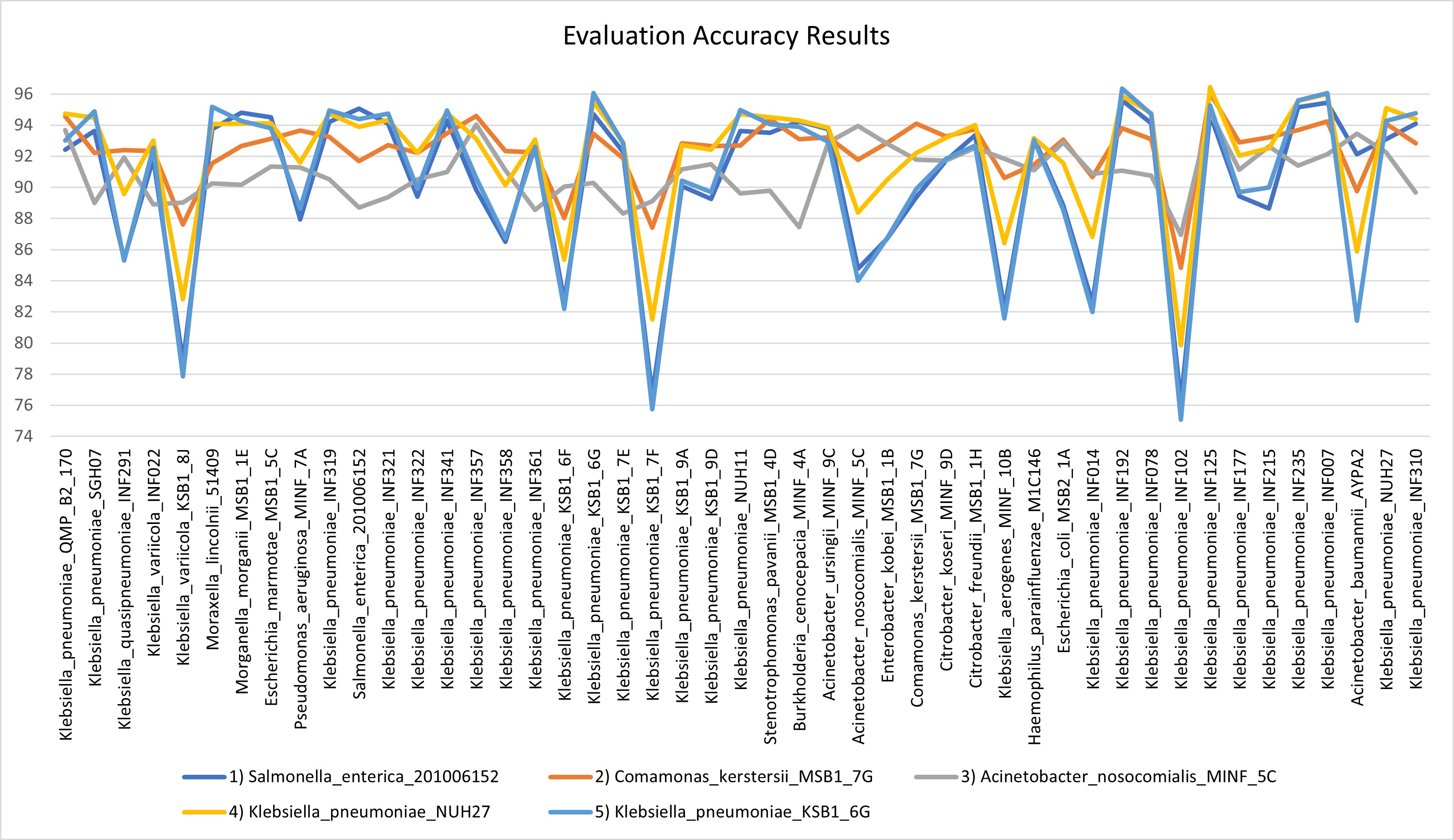}
    \caption{Training-setup accuracy results}
    \label{fig:acc_eval_res}
\end{figure}

First, a plot is illustrated in Fig. \ref{fig:acc_eval_res} showing accuracy improvement after each stage on various bacterial data. The first stage pointed out with the blue line indicating "Salmonella eneterica 201006152" bacteria has the lowest accuracy value in most cases. This model has an accuracy below 80\% for three bacterial species. Subsequently, transfer learning applied to "Comamonas kerstersii MSB1 7G" data shows improvement in most bacterial samples. Also, this model has improved the accuracy in mentioned three points on the graph, and it has an accuracy lower than 90\% for only five bacterial data. Third, results illustrate a significant drop when training on "Acinetobacter nosocomialis MINF 5C" as the results show accuracy higher than 91\% for only 26 bacterial samples.

In contrast, 41 bacterial samples with higher accuracy than 91\% for the model trained on "Comamonas kerstersii MSB1 7G" data. Next, the transfer learning model on "Klebsiella pneumoniae NUH27" shows improvement with having an accuracy of more than 94\% for 20 out of 50 bacterial samples. Subsequently, we have used "Klebsiella pneumoniae KSB1 6G" bacterial data to perform the transfer learning that has resulted in a model behavior similar to the model first trained on "Salmonella eneterica 201006152" bacteria. 

Among these trained models, the top two models with maximum average accuracy across are models trained at the second and fourth stage (on "Comamonas kerstersii MSB1 7G" and "Klebsiella pneumoniae NUH27" data) with accuracies respectively 92.84\% and 93.16\%. In the next section, we investigate the model out of these two with the highest accuracy for inference (experimental setup), as using the model in that stage with a greedy setup leads to an accuracy drop.

\subsubsection{Inference accuracy evaluation}

In the inference stage, we have two options to connect the feedback part from the output of the LRDB to its decoder input. First, use a greedy search method that feeds the predicted DNA value in each iteration as the decoder's input for the next step. This method decreases the accuracy as no further analysis is performed on the output, but it does not reduce the model speed. Second, the Beam search method calculates and tracks k best outputs during the inference stage and improves accuracy. On the other hand, this method takes more time for the base calling algorithm to be executed.

We have used the greedy search method in our simulations. However, it is noteworthy to state that the accuracy results can be further improved using Beam search methods, and also, the user can set up their desired feedback mechanism based on the experimental constraints. For instance, they can use Beam search if the timing is not essential or use the greedy search in time-critical applications.

\begin{figure}
    \centering
    \includegraphics[width=\columnwidth]{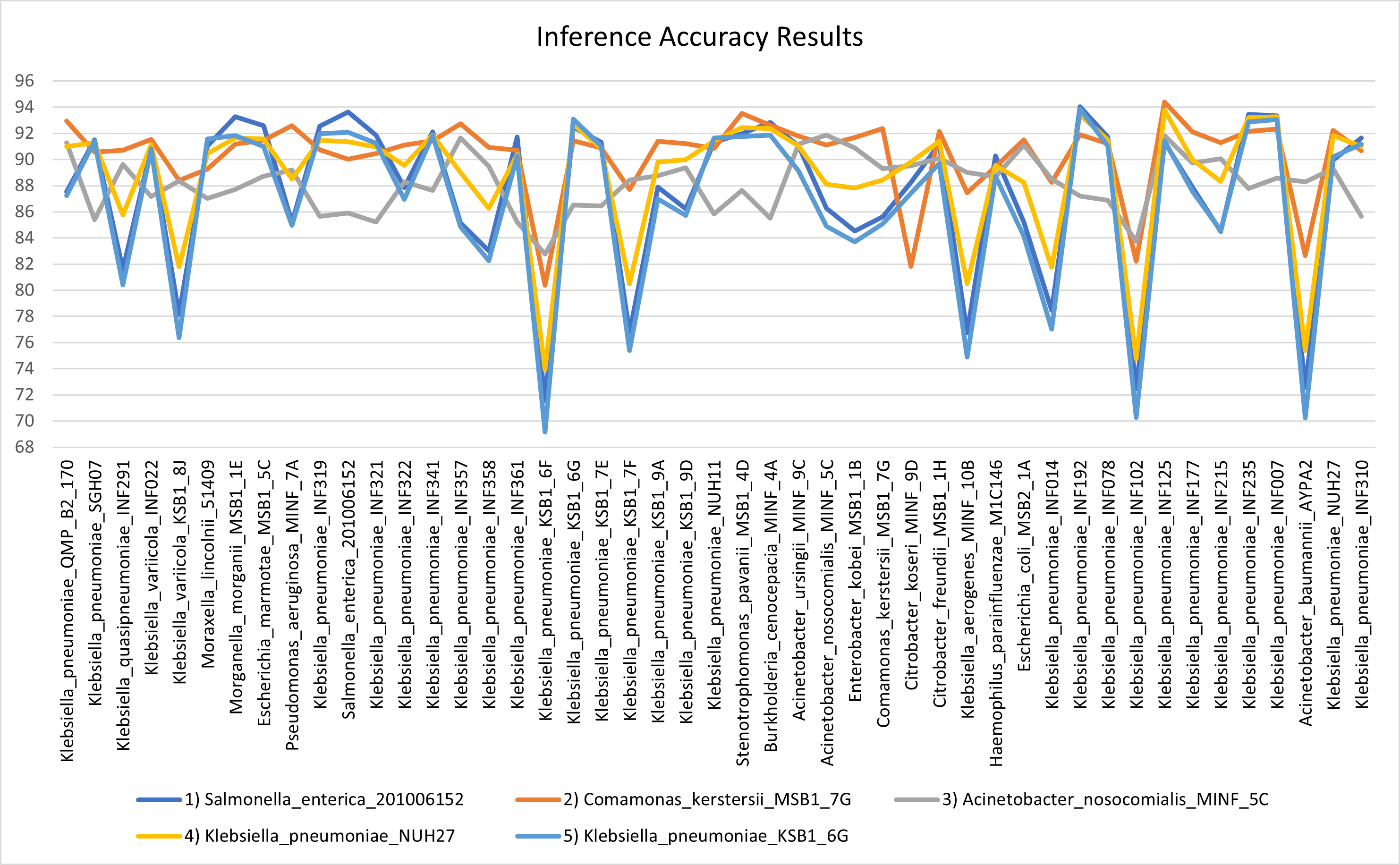}
    \caption{Inference-setup accuracy results plotted for bacterial data}
    \label{fig:acc_inf_res}
\end{figure}

\begin{figure}
    \centering
    \includegraphics[width=\columnwidth]{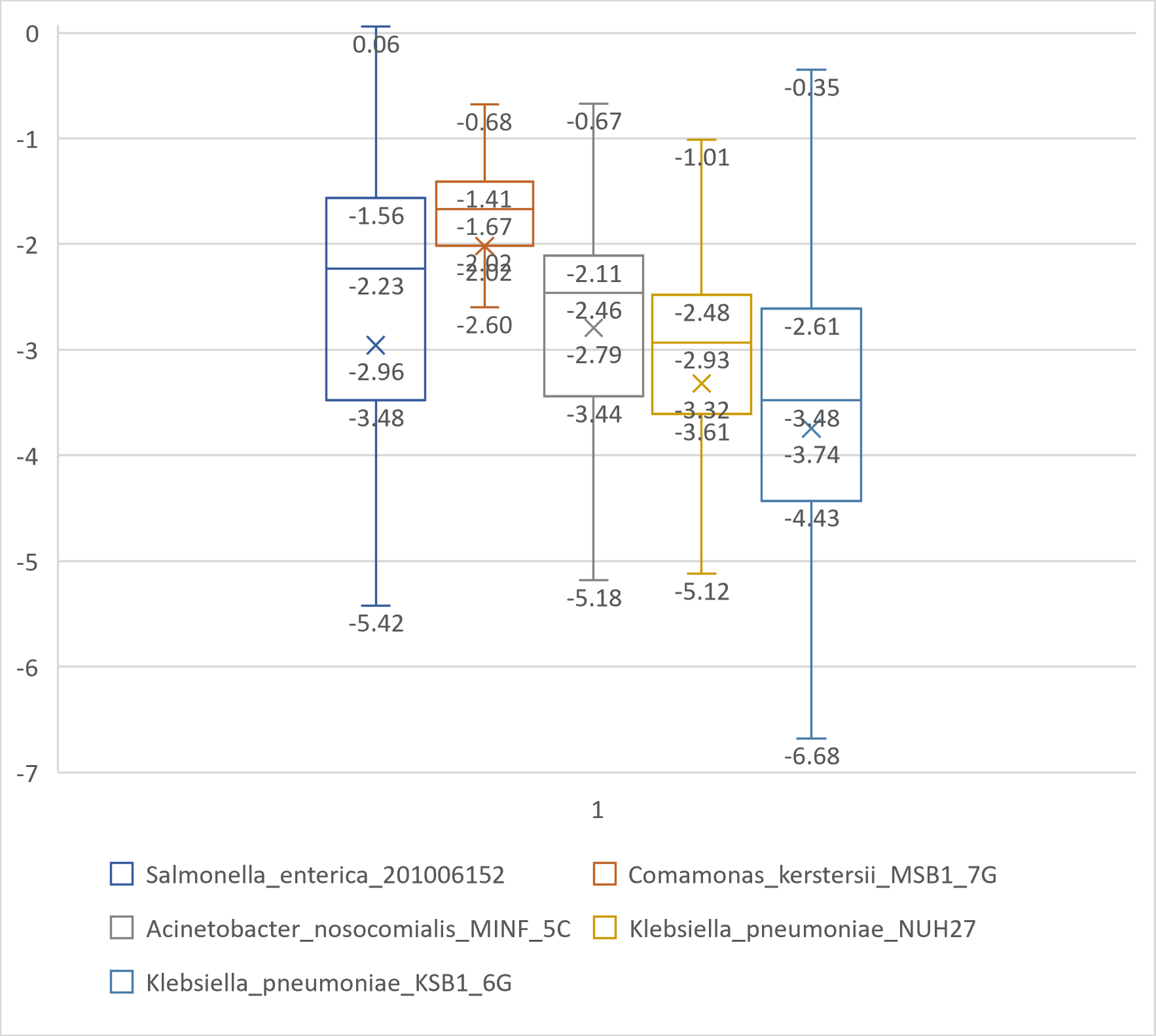}
    \caption{Accuracy drop measured using greedy search method (The accuracy of the model in the inference stage is subtracted from evaluation accuracy of the model during the training phase using the same evaluation bacterial data)}
    \label{fig:acc_drop}
\end{figure}

The same models from section \ref{sec:tlaae} are set up in a greedy manner, and their inference accuracy results on the test bacterial samples are shown in Fig. \ref{fig:acc_inf_res}. The results also confirm our discussion on the accuracy drop using the greedy search. Among various models, based on Fig. \ref{fig:acc_drop} the model trained at the second stage  (using "Comamonas kerstersii MSB1 7G" data)  indicates the most negligible accuracy drop. 

\begin{figure*}[ht]
    \centering
    \includegraphics[width=1.3\columnwidth]{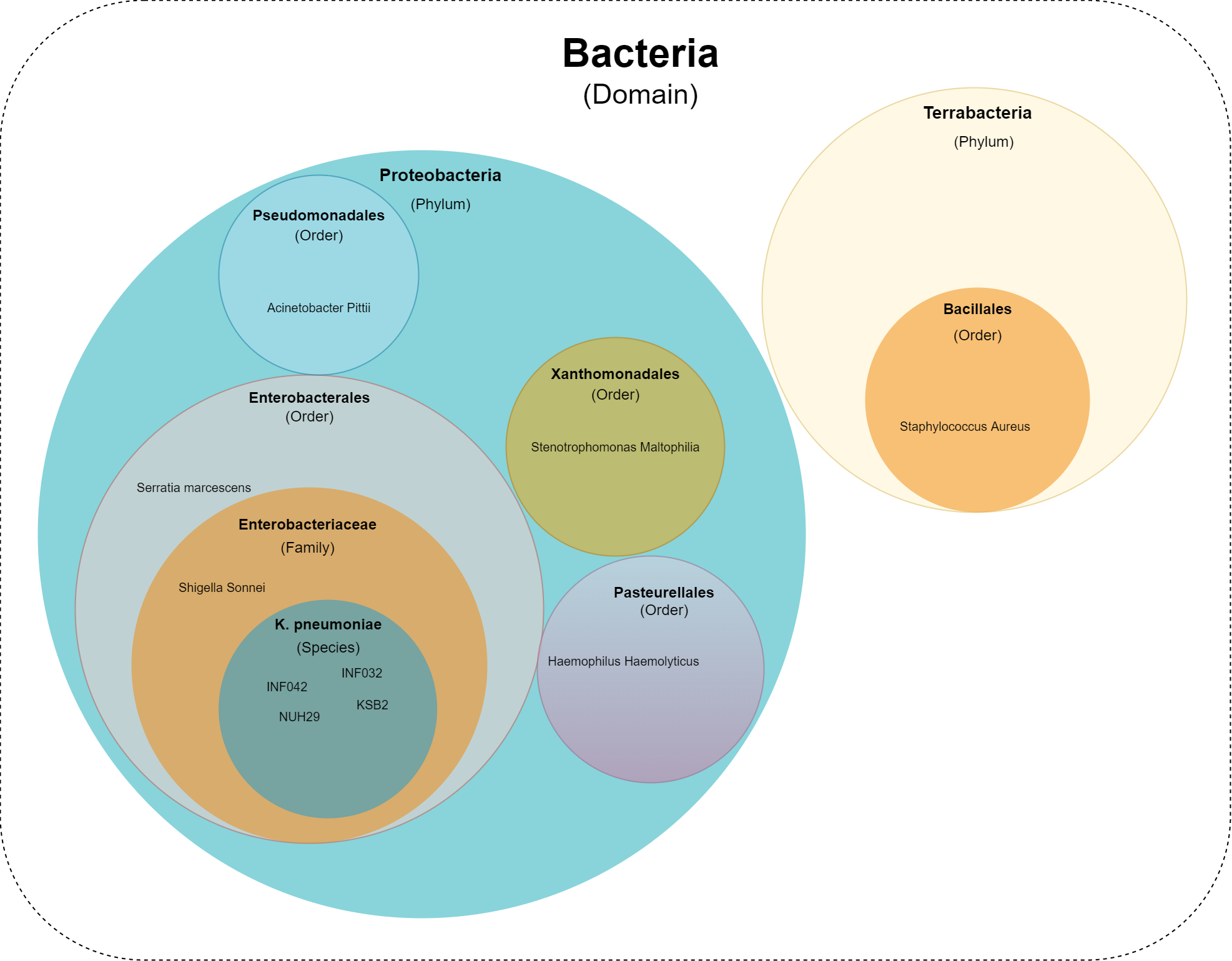}
    \caption{The hierarchical illustration of different bacterial species based on their phylum, order, family, and species.}
    \label{fig:bac_samp}
\end{figure*}

For evaluating the read identity in the next section, we have chosen the model trained in the second stage ("Comamonas kerstersii MSB1 7G" data) because it has the lowest accuracy drop, makes it the model with the highest inference accuracy in comparison with a model trained on "Klebsiella pneumoniae NUH27". In more detail, a model trained on "Klebsiella pneumoniae NUH27" has an average accuracy of 93.16\% and a drop of -3.32\% leading to an average accuracy of 89.84\%. In contrast, the model for "Comamonas kerstersii MSB1 7G" has an average accuracy of 92.84\% and a drop of 1.67\%, making it the model with an average accuracy of 91.17\%.

\subsection{Read identity and Accuracy Comparison}
\label{sec:perf_metrics}
The most critical performance metric of a base caller is the real identity. We have a base called ten different bacterial data without having their DNA strands \cite{Wick2019_raw}, and then, we have used the base calling results using Blast package to measure the read identity following 'BLAST identity' definition (\href{https://lh3.github.io/2018/11/25/on-the-definition-of-sequence-identity}{https://lh3.github.io/2018/11/25/on-the-definition-of-sequence-identity}). The reference genomes are also provided in \cite{Wick2019_ref}. Fig. \ref{fig:bac_samp} illustrates a diagram with each circle representing the domain, phylum, classes, and orders for our inference of bacterial samples. The target for measuring the real identity is Enterobacterales and Pseudomonadales bacterial orders because the base callers are trained to base call in these orders. 

We have a base called all these bacterial samples using our LRDB, which is trained on "Salmonella enterica 201006152" and fine-tuned with "Comamonas kerstersii MSB1 7G" and "Acinetobacter nosocomialis MINF 5C" bacterial data. Next, the median read identities corresponding to each bacterial sample are shown in Table \ref{tab:read_identity}. 

The data in the Table indicates that the LRDB achieves a high read identity for both bacterial phylums of test data. Moreover, "Acinetobacter Pitti" has achieved the highest read identity with 94.623\% as the LRDB is trained on a closely related bacteria ("Acinetobacter nosocomialis MINF 5C") from the same order. Next, the bacterias in the Enterobacterales order have achieved a high read identity as the LRDB is trained on the same order with the "Salmonella enterica 201006152" sample. The average read identity in this order for all bacterial species is higher than other base callers, with an average of 91.569\%.

On the other hand, as expected "Haemophilus Haemolyticus" and "Stenotrophomonas Maltophilia" did not achieve a desirable read identity. This is because these species are from Xanthomonadales and Pasteurellales orders, respectively, which our LRDB has not seen during the training procedure. The user can further train the LRDB for both species and reach reasonable read identities, as our active learning approach explains. Moreover, the LRDB achieves read identity values higher than 90\% in 8 out of 10 samples, and the read identity average reported in the Table for all bacteria species except "Haemophilus Haemolyticus" and "Stenotrophomonas Maltophilia" show 0.283\% read identity improvement. The importance of this improvement is highlighted as we have only used three bacterial species to train our LRDB compared to other models, which are trained on an entire database consisting of 50 bacterial species.

\begin{figure}
    \centering
    \includegraphics[width=\columnwidth]{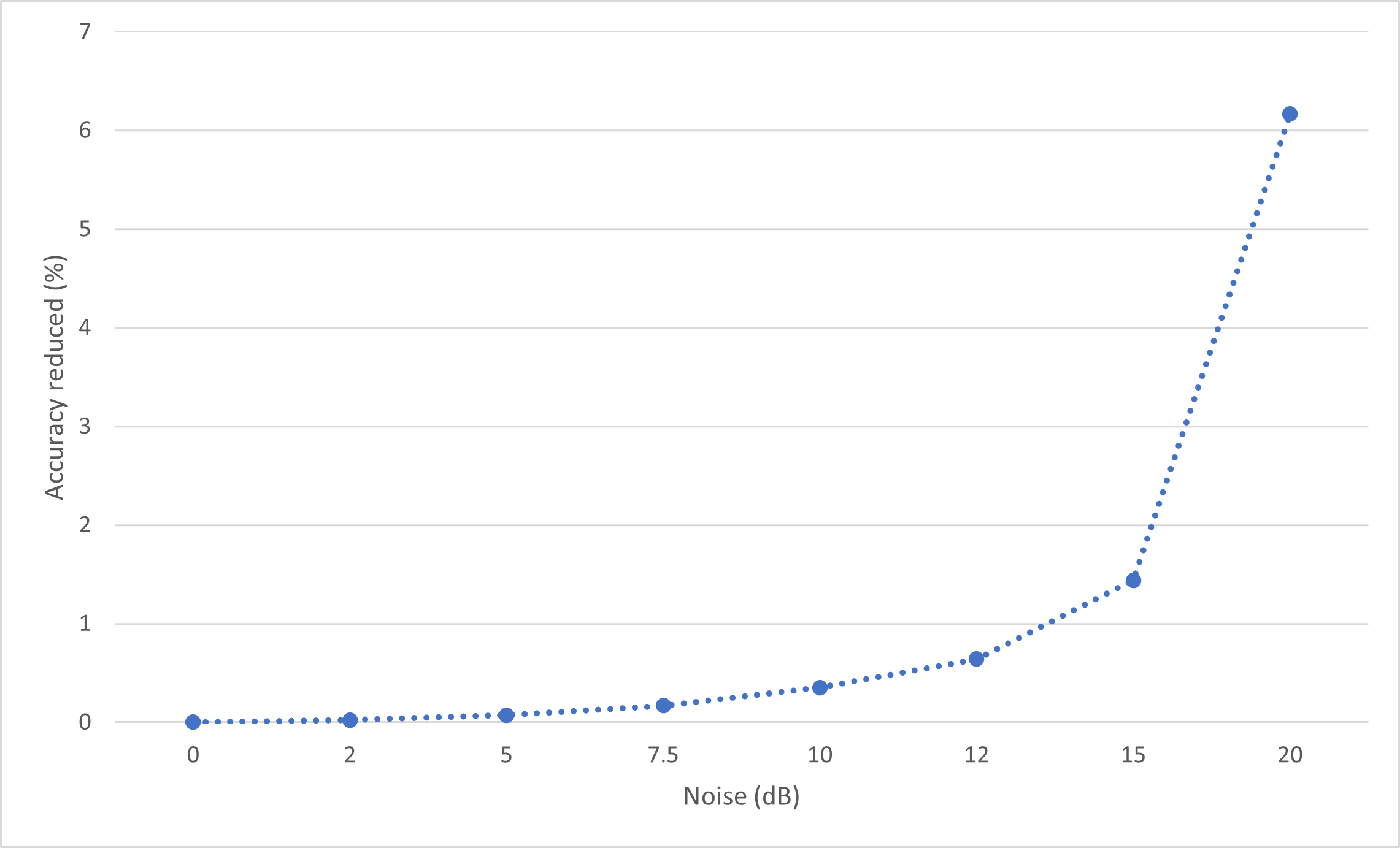}
    \caption{Percentage of accuracy reduction per noise injected in dB}
    \label{fig:noise_red}
\end{figure}

\begin{table*}
    \centering
    \caption{Different bacterial samples' read identity for using various base callers (Median of real identity is reported in this Table)}
    \label{tab:read_identity}
    \begin{tabular}{lccccccc}
    \toprule
        Bacterial Sample & LRDB & CATCaller$^{f32}$ & SACall & Guppy-KP & Guppy & Albacore\\ \midrule
        Klebsiella Pneumoniae NUH29 & 91.252 & 91.511 & 91.243 & 89.384 & 89.468 & 87.105\\
        Klebsiella Pneumoniae KSB2 & 91.192 & 90.974 & 90.583 & 88.229 & 89.009 & 86.548 \\
        Klebsiella Pneumoniae INF042 & 91.979 & 91.181 & 90.852 & 88.510 & 89.399 & 86.548 \\
        Klebsiella Pneumoniae INF032 & 92.285 & -  & - & 88.79 & 87.14 & -\\
        Shigella Sonnei & 92.599 & 91.247 & 90.787 & 88.346 & 90.628 & 88.015 \\
        Serratia Marcescens & 90.109 & 91.156  & 90.917& 88.615 & 91.120 & 87.053\\
        Haemophilus Haemolyticus & 58.380 & 92.614 & 92.308 & 89.697 & 92.233 & 88.502\\
        Stenotrophomonas Maltophilia & 83.233 & 90.704 & 90.507 & 88.741 & 89.393 & 87.195 \\
        Acinetobacter Pittii & 94.623 & 91.324 & 90.890 & 88.623 & 92.354 & 87.995 \\
        Staphylococcus Aureus & 90.081 & 92.984 & 91.692 & 90.692 & 94.638 & 90.989 \\ \midrule
        Average on Enterobacterales order & 91.569 & 91.213 & 90.876 & 88.645 & 89.460 & 87.053 \\
        Average on generalized bacteria & 91.765 & 91.482 & 90.994& 88.898& 90.469& 87.750 \\
        
        \bottomrule
    \end{tabular}

\end{table*}

\subsection{Noise results}

In the meantime, when Min-ION devices are producing the signal sequences, environmental noise is added to the measurements. Such noise stems from the environment white noise, measurement fault, and electrical noise. Hence, base callers should be noise-tolerant, so their performance does not degrade in real-world scenarios. Worst-case initial SNR is about 2 dB in such environments \cite{shekar2019wavelet}; however, we have simulated a more extensive range of probable noise values to see if our network is tolerant against these values.

To create the noisy signal data, we have used various powers of white noise in our model: ranging from 2 dB to 20 dB. Then, we used the inference setup and calculated the accuracy reduction for various noise signal data of bacterial samples. Finally, we have computed the average noise reduction throughout our samples and reported them in Fig. \ref{fig:noise_red}. As observed in this figure, the accuracy reduction is lower than 1.5\% when injecting 15 dB noise to input data (5 times the normal environmental noise); we observe 0.024\% and 0.073\% accuracy reduction for 2dB and 5dB noise, respectively. These results justify the noise tolerance of the LSTM caller.

\subsection{Parameter and speed report}
As demonstrated in section \ref{sec:param_setting}, various set-ups of LRDB can be used for base calling. Each of these models varies in performance as the lightweight models perform faster and have fewer model parameters so that the user can select the proper model based on his/her system configuration. This selection from table \ref{tab:model_params} can affect base calling results by a 2\% accuracy drop if we only consider models with 100,000 parameters.

Another approach applicable to LRDB is enhancing speed through parallelism in the encoder and decoder parts, while both are connected in the attention mechanism, and the decoder runs independently from the encoder. To parallelize the network, we run the encoder once and store the outputs. Next, we use this stored output each time we want to run the inference stage for base calling.

\begin{figure}
    \centering
    \includegraphics[width=\columnwidth]{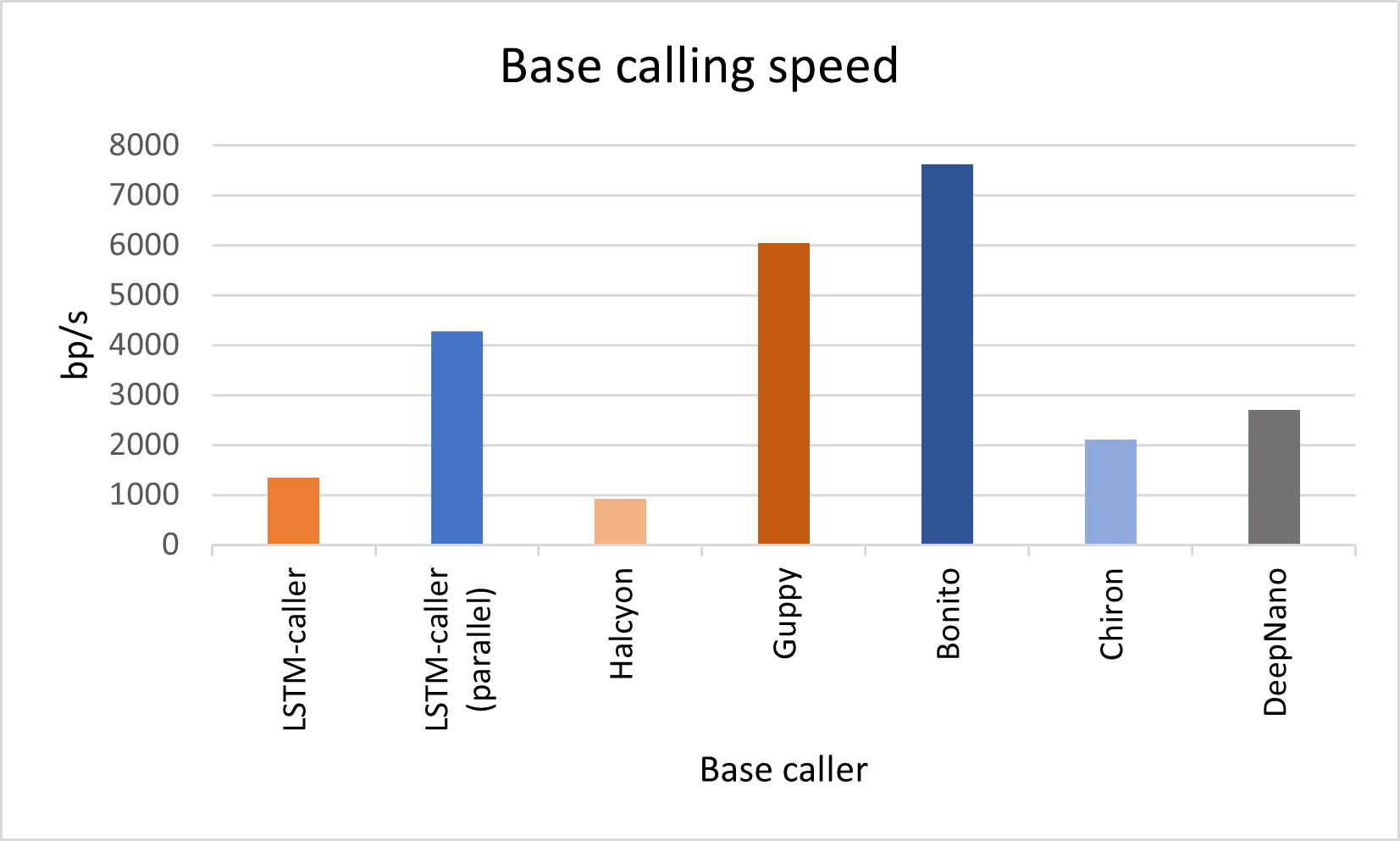}
    \caption{Speed performance report for various base callers (LRDB (parallel) stands for parallel execution of LRDB).}
    \label{fig:bsc_spd}
\end{figure}

This parallelism leads to a high speed up as the encoder part contains most of the model parameters to be computed, and we avoid these redundant computations each time. The simulations demonstrate that even before the parallelism, LRDB with 1,347 bp/s satisfies the input speed rate of data from a Min-ION device (450 bp/s) \cite{abbaszadegan2019encoder} and it shows that it can operate as fast as 4,284 bp/s by applying the parallel execution. Although our setup cannot be compared with the GPU setup in the other papers, we have reported our speed results next to the results in \cite{konishi2021halcyon} in Fig. \ref{fig:bsc_spd}. The results show that LRDB performs in the desired speed range as all these base callers are for practical usage.

\section{Conclusion and Future work}
In this paper, we elaborated on several aspects of base callers' active usage, which are not covered in prior research works. Due to privacy issues, it is of value for researchers and biologists to train and use their models. In this case, known open-source models help maintain privacy and to shorten the training process, we should take measures against the naive usage of data in the training stage. In the LRDB, we benefit from a transfer learning technique to reach the desirable generalization using only five bacterial samples instead of training the model over 50 provided samples \cite{wick2019performance}. Additionally, the variety of data (50 bacterial samples) used in prior works is excessive for a light Deep Learning model to be trained. The model cannot simply learn accurate information about all the training data. This further leads to a well-known problem called catastrophic forgetting, which we have solved by using transfer learning and fine-tuning as two standard techniques for such issues with a constraint on the train data volume. Also, LRDB can be trained and extended for other bacterial samples only by fine-tuning the existing model, which is helpful by shortening the time-consuming training phase for base callers being trained from scratch.

During our detailed paper, we have proposed a lightweight LRDB model for base calling purposes, with increased read identity for predefined target bacterial samples. In the case of generalized and target bacterial samples, we have respectively 0.28\% and 0.35\% read identity improvement over cutting-edge base callers. The noise-tolerant properties of the LRDB show that it suffers from a 1.439\% accuracy loss for 15dB noise injection (5 times the normal environmental noise). Finally, we have reported the performance metrics of the selected model showing that the model can be used in real-world scenarios.

\bibliography{document}

\begin{thebibliography}{10}
\expandafter\ifx\csname url\endcsname\relax
  \def\url#1{\texttt{#1}}\fi
\expandafter\ifx\csname urlprefix\endcsname\relax\def\urlprefix{URL }\fi
\expandafter\ifx\csname href\endcsname\relax
  \def\href#1#2{#2} \def\path#1{#1}\fi

\bibitem{howorka2001sequence}
S.~Howorka, S.~Cheley, H.~Bayley, Sequence-specific detection of individual dna
  strands using engineered nanopores, Nature biotechnology 19~(7) (2001)
  636--639.

\bibitem{taheri2021high}
M.~Taheri, H.~Zandevakili, A.~Mahani, A high-performance memristor-based
  smith-waterman dna sequence alignment using fpni structure, Journal of
  Applied Research in Electrical Engineering 1~(1) (2021) 59--68.

\bibitem{taheridevelopment}
M.~Taheri, A.~Mahani, Development and hardware acceleration of a novel 2-d
  bwa-mem dna sequencing alignment algorithm, 1st Conference on Applied
  Research in Electrical Engineering (AREE) (2021).

\bibitem{nan_spd}
B.~CG.,
  \href{https://nanoporetechcom/resource-centre/videos/wafer-thin-update}{Oxford
  nanopore technologies: a wafer thin update.}, . (2016).
\newline\urlprefix\url{https://nanoporetechcom/resource-centre/videos/wafer-thin-update}

\bibitem{taheri2021novel}
M.~Taheri, A.~Mahani, A novel 2-d bwa-mem fpga accelerator for short-read
  mapping of the whole human genome, Journal of Applied Research in Electrical
  Engineering 1~(2) (2021) 203--210.

\bibitem{bovza2017deepnano}
V.~Bo{\v{z}}a, B.~Brejov{\'a}, T.~Vina{\v{r}}, Deepnano: deep recurrent neural
  networks for base calling in minion nanopore reads, PloS one 12~(6) (2017)
  e0178751.

\bibitem{david2017nanocall}
M.~David, L.~J. Dursi, D.~Yao, P.~C. Boutros, J.~T. Simpson, Nanocall: an open
  source basecaller for oxford nanopore sequencing data, Bioinformatics 33~(1)
  (2017) 49--55.

\bibitem{magi2018nanopore}
A.~Magi, R.~Semeraro, A.~Mingrino, B.~Giusti, R.~D’Aurizio, Nanopore
  sequencing data analysis: state of the art, applications and challenges,
  Briefings in bioinformatics 19~(6) (2018) 1256--1272.

\bibitem{stoiber2017basecrawller}
M.~Stoiber, J.~Brown, Basecrawller: streaming nanopore basecalling directly
  from raw signal, BioRxiv (2017) 133058.

\bibitem{teng2018chiron}
H.~Teng, M.~D. Cao, M.~B. Hall, T.~Duarte, S.~Wang, L.~J. Coin, Chiron:
  translating nanopore raw signal directly into nucleotide sequence using deep
  learning, GigaScience 7~(5) (2018) giy037.

\bibitem{wick2019performance}
R.~R. Wick, L.~M. Judd, K.~E. Holt, Performance of neural network basecalling
  tools for oxford nanopore sequencing, Genome biology 20~(1) (2019) 1--10.

\bibitem{huang2020sacall}
N.~Huang, F.~Nie, P.~Ni, F.~Luo, J.~Wang, Sacall: a neural network basecaller
  for oxford nanopore sequencing data based on self-attention mechanism,
  IEEE/ACM Transactions on Computational Biology and Bioinformatics (2020).

\bibitem{zeng2020causalcall}
J.~Zeng, H.~Cai, H.~Peng, H.~Wang, Y.~Zhang, T.~Akutsu, Causalcall: Nanopore
  basecalling using a temporal convolutional network, Frontiers in genetics 10
  (2020) 1332.

\bibitem{lv2020end}
X.~Lv, Z.~Chen, Y.~Lu, Y.~Yang, An end-to-end oxford nanopore basecaller using
  convolution-augmented transformer, in: 2020 IEEE International Conference on
  Bioinformatics and Biomedicine (BIBM), IEEE, 2020, pp. 337--342.

\bibitem{konishi2021halcyon}
H.~Konishi, R.~Yamaguchi, K.~Yamaguchi, Y.~Furukawa, S.~Imoto, Halcyon: an
  accurate basecaller exploiting an encoder--decoder model with monotonic
  attention, Bioinformatics 37~(9) (2021) 1211--1217.

\bibitem{torrey2010transfer}
L.~Torrey, J.~Shavlik, Transfer learning, in: Handbook of research on machine
  learning applications and trends: algorithms, methods, and techniques, IGI
  global, 2010, pp. 242--264.

\bibitem{howard2018universal}
J.~Howard, S.~Ruder, Universal language model fine-tuning for text
  classification, arXiv preprint arXiv:1801.06146 (2018).

\bibitem{huang2019attention}
N.~Huang, F.~Nie, P.~Ni, F.~Luo, J.~Wang, An attention-based neural network
  basecaller for oxford nanopore sequencing data, in: 2019 IEEE International
  Conference on Bioinformatics and Biomedicine (BIBM), IEEE, 2019, pp.
  390--394.

\bibitem{johnson2017real}
S.~S. Johnson, E.~Zaikova, D.~S. Goerlitz, Y.~Bai, S.~W. Tighe, Real-time dna
  sequencing in the antarctic dry valleys using the oxford nanopore sequencer,
  Journal of Biomolecular Techniques: JBT 28~(1) (2017) 2.

\bibitem{edwards2017deep}
A.~Edwards, A.~Soares, S.~M. Rassner, P.~Green, J.~Felix, A.~C. Mitchell, Deep
  sequencing: intra-terrestrial metagenomics illustrates the potential of
  off-grid nanopore dna sequencing, BioRxiv (2017) 133413.

\bibitem{edwards2016extreme}
A.~Edwards, A.~R. Debbonaire, B.~Sattler, L.~Mur, A.~J. Hodson, Extreme
  metagenomics using nanopore dna sequencing: a field report from svalbard, 78
  n, BioRxiv 10 (2016) 073965.

\bibitem{castro2017nanopore}
S.~L. Castro-Wallace, C.~Y. Chiu, K.~K. John, S.~E. Stahl, K.~H. Rubins, A.~B.
  McIntyre, J.~P. Dworkin, M.~L. Lupisella, D.~J. Smith, D.~J. Botkin, et~al.,
  Nanopore dna sequencing and genome assembly on the international space
  station, Scientific reports 7~(1) (2017) 1--12.

\bibitem{taheri2021hardware}
M.~Taheri, M.~S. Ansari, S.~Magierowski, A.~Mahani, Hardware acceleration of
  the novel two dimensional burrows-wheeler aligner algorithm with maximal
  exact matches seed extension kernel, IET Circuits, Devices \& Systems 15~(2)
  (2021) 94--103.

\bibitem{mitsuhashi2017portable}
S.~Mitsuhashi, K.~Kryukov, S.~Nakagawa, J.~S. Takeuchi, Y.~Shiraishi, K.~Asano,
  T.~Imanishi, A portable system for rapid bacterial composition analysis using
  a nanopore-based sequencer and laptop computer, Scientific reports 7~(1)
  (2017) 1--9.

\bibitem{quick2015rapid}
J.~Quick, P.~Ashton, S.~Calus, C.~Chatt, S.~Gossain, J.~Hawker, S.~Nair,
  K.~Neal, K.~Nye, T.~Peters, et~al., Rapid draft sequencing and real-time
  nanopore sequencing in a hospital outbreak of salmonella, Genome biology
  16~(1) (2015) 1--14.

\bibitem{greff2016lstm}
K.~Greff, R.~K. Srivastava, J.~Koutn{\'\i}k, B.~R. Steunebrink, J.~Schmidhuber,
  Lstm: A search space odyssey, IEEE transactions on neural networks and
  learning systems 28~(10) (2016) 2222--2232.

\bibitem{pu2016variational}
Y.~Pu, Z.~Gan, R.~Henao, X.~Yuan, C.~Li, A.~Stevens, L.~Carin, Variational
  autoencoder for deep learning of images, labels and captions, Advances in
  neural information processing systems 29 (2016) 2352--2360.

\bibitem{srivastava2015unsupervised}
N.~Srivastava, E.~Mansimov, R.~Salakhudinov, Unsupervised learning of video
  representations using lstms, in: International conference on machine
  learning, PMLR, 2015, pp. 843--852.

\bibitem{casjens1998diverse}
S.~Casjens, The diverse and dynamic structure of bacterial genomes, Annual
  review of genetics 32~(1) (1998) 339--377.

\bibitem{Wick2019_train}
R.~Wick,
  \href{https://bridges.monash.edu/articles/dataset/Training_data/76761984}{Training
  data, 10.26180/5c5a5f5ff20ed}, . (2 2019).
\newblock \href {https://doi.org/10.26180/5c5a5f5ff20ed}
  {\path{doi:10.26180/5c5a5f5ff20ed}}.
\newline\urlprefix\url{https://bridges.monash.edu/articles/dataset/Training_data/76761984}

\bibitem{Wick2019_raw}
R.~Wick,
  \href{https://bridges.monash.edu/articles/dataset/Raw_fast5s/7676174}{Raw
  fast5s, 10.26180/5c5a5fa08bbee}, . (5 2019).
\newblock \href {https://doi.org/10.26180/5c5a5fa08bbee}
  {\path{doi:10.26180/5c5a5fa08bbee}}.
\newline\urlprefix\url{https://bridges.monash.edu/articles/dataset/Raw_fast5s/7676174}

\bibitem{Wick2019_ref}
R.~Wick,
  \href{https://bridges.monash.edu/articles/dataset/Reference_genomes/7676135}{Reference
  genomes, 10.26180/5c5a5fcf72e40}, . (2 2019).
\newblock \href {https://doi.org/10.26180/5c5a5fcf72e40}
  {\path{doi:10.26180/5c5a5fcf72e40}}.
\newline\urlprefix\url{https://bridges.monash.edu/articles/dataset/Reference_genomes/7676135}

\bibitem{shekar2019wavelet}
S.~Shekar, C.-C. Chien, A.~Hartel, P.~Ong, O.~B. Clarke, A.~Marks, M.~Drndic,
  K.~L. Shepard, Wavelet denoising of high-bandwidth nanopore and ion-channel
  signals, Nano letters 19~(2) (2019) 1090--1097.

\bibitem{abbaszadegan2019encoder}
M.~Abbaszadegan, An encoder-decoder based basecaller for nanopore dna
  sequencing, Master’s Thesis, York University, Toronto, ON, Canada (2019).

\end{thebibliography}

\end{document}